\documentclass[12pt,twoside]{article}

\usepackage{amsfonts}
\usepackage{amsmath}
\usepackage{cite}
\usepackage{epsfig}
\usepackage{a4}

\newcommand{\I}{\mathrm{i}}        
\newcommand{\E}{\mathrm{e}}        
\newcommand{\D}{\mathrm{d}}       

\newcommand{\sign}{\operatorname{sign}} 

\newcommand{\Det}{\operatorname{det}}
\newcommand{\perm}{\operatorname{perm}}

\begin{document}

\thispagestyle{empty}

\begin{center}

{\Large {\bf From multiple integrals to Fredholm determinants\\}}

\vspace{7mm}

{\large Alexander Seel\footnote[1]{e-mail:
 alexander.seel@itp.uni-hannover.de}, 
Frank G\"ohmann\footnote[2]{e-mail: goehmann@physik.uni-wuppertal.de}\\
and Andreas Kl\"{u}mper\footnote[3]{e-mail: kluemper@physik.uni-wuppertal.de}\\

\vspace{5mm}

$^1$Institut f\"ur Theoretische Physik, Leibniz Universit\"at Hannover,\\
30167 Hannover, Germany\\[2ex]

$^{2,3}$Fachbereich C -- Physik, Bergische Universit\"at Wuppertal,\\
42097 Wuppertal, Germany\\}

\vspace{20mm}

{\large {\bf Abstract}}

\end{center}

\begin{list}{}{\addtolength{\rightmargin}{10mm}
               \addtolength{\topsep}{-5mm}}
\item
We consider a multiple integral representation for the
finite temperature density-density correlation functions
of the one-dimensional Bose gas with delta function
interaction in the limits of infinite and vanishing repulsion.
In the former case a known Fredholm determinant is recovered.
In the latter case a similar expression appears with permanents
replacing determinants.\\[2ex]

{\it PACS: 02.30.Ik, 05.30.-d, 05.30.Jp}
\end{list}

\clearpage

\section{Introduction}
The one-dimensional Bose gas with delta function interaction (contact
interaction) is a paradigmatic solvable \cite{LiLi63} many-body
quantum system. It shares its $R$-matrix underlying the integrability
with the spin-$\frac12$ Heisenberg chain, but may be considered even
simpler, since its Bethe ansatz equations have only real roots
\cite{KBIBo}. This feature, the absence of strings, in first place
simplifies the thermodynamic Bethe ansatz (TBA) analysis, which is
the reason why the Bose gas was the first non-trivial Bethe ansatz
solvable model whose thermodynamics was analyzed in detail \cite{YaYa69}.

When in the 80s and early 90s the algebraic Bethe ansatz
was applied to the study of correlation functions of solvable
models, the Bose gas was again in the center of interest \cite{KBIBo}.
Explicit results, generalizing early work \cite{Schultz63,Lenard64,%
Lenard66}, were obtained in particular for the case of infinite
repulsion, the so called impenetrable Boson limit. In this limit
many correlation functions can be expressed as Fredholm determinants
or Fredholm minors related to integral operators of a special type,
the so-called integrable integral operators \cite{IIK90a,KoSl90,KBIBo},
which have a close connection with classical integrable evolution
equations and for this reason are suitable for an explicit calculation of
the asymptotics of correlation functions.

Another type of expression for the correlation functions of solvable
many-body systems are the so-called multiple integral representations
first obtained in \cite{JMMN92} for the density matrix of a segment
of the XXZ Heisenberg chain by an approach based on the representation
theory of the quantum affine algebra $U_q (\widehat{\mathfrak sl}_2)$.
Later this result was rederived by means of the algebraic Bethe ansatz
\cite{KMT99b} which was important, since the algebraic Bethe ansatz
turned out to be flexible enough for a number of generalizations.
In our context we would like to mention the works \cite{KMST02a} and
\cite{GoKlSe04}, where multiple integral representations for a
one-parameter generating function of the $zz$-correlation functions of
the XXZ chain were derived for the ground state and for finite
temperature. From \cite{KMST02a} it was only a small step to the
corresponding expression for the Bose gas \cite{KKMST07}, namely a
multiple integral representation for a generating function of the
density-density correlation functions. A finite temperature version was
obtained directly from \cite{GoKlSe04} in a certain scaling limit
close to the ferromagnetic point $\Delta = -1$ of the XXZ chain
\cite{SBGK07a}.

The latter is the starting point of this work and is reviewed
in the next section. Then, in section \ref{sechardcore}, we
perform the impenetrable Boson limit and show that the resulting
expression is equal to a known Fredholm determinant representation.
This is the main result we wish to convey: in an appropriate limit
the multiple integrals turn into a Fredholm determinant. Looking
at it from a different angle this means, that we may interpret the
multiple integrals as a `deformation of a Fredholm determinant'.
In section \ref{secsoftcore} we complement our work with an account
of the free Boson limit which is less trivial as might appear at
first sight. In fact we derive a multiple integral formula with
permanents replacing the determinants of the impenetrable case, which
we were not able to spot in the literature.
To round off this work we demonstrate in section \ref{sec12PF} how to
obtain the density-density correlation function of the Bose gas
from the generating function in the two limiting cases of free and
of impenetrable Bosons. Section \ref{secconcl} is devoted to a
concluding summary.


\section{Density-Density Correlations in the Bose Gas} \label{secDDCBG}
The one-dimensional Bose gas with contact interaction is described
by the Hamiltonian
\begin{equation}
     H = \int_0^\ell \D z \,
         \big[ (\partial_z\psi^\dagger) (\partial_z \psi)
	 + c \, \psi^\dagger(z)\psi^\dagger(z)\psi(z)\psi(z)\big] \, .
\end{equation}
Here $\psi^\dagger(z)$ and $\psi(z)$ are Bose fields with canonical
equal-time commutation relations which act on the interval $[0, \ell]$
for which we assume periodic boundary conditions. $c > 0$ is the
coupling constant.

The operator
\begin{equation}
     Q (x) = \int_0^x \D z \,\psi^\dagger(z)\psi(z)
\end{equation}
measures the number of particles in the interval $[0,x]$,
$0 \le x \le \ell$. We need it to define a one-parameter generating
function of the density-density correlation functions by
$\langle \E^{\varphi Q (x)} \rangle_{T, \mu}$, where the brackets
indicate the grand-canonical ensemble average for a heat and particle
bath of temperature $T$ and with chemical potential $\mu$. The function
$\langle \E^{\varphi Q (x)} \rangle_{T, \mu}$ is particularly convenient
in the context of the algebraic Bethe ansatz \cite{KBIBo}. With the
shorthand notation $j (x) = \psi^\dagger(x) \psi(x)$ for the particle
density operator we have the following formulae,
\begin{equation} \label{applgenfun}
     \langle j(x)\rangle_{T, \mu} =
        \partial_x \partial_\varphi
	  \langle \textrm{e}^{\varphi Q (x)} \rangle_{T, \mu}
	     \Bigr|_{\varphi=0} \, , \quad 
	  \langle j(0) j(x) \rangle_{T, \mu}
        = \frac12 \partial^2_x \partial^2_\varphi
	  \langle \textrm{e}^{\varphi Q (x)}\rangle_{T, \mu}
	     \Bigr|_{\varphi=0} \, .
\end{equation}

In our previous work \cite{SBGK07a} we obtained the multiple
integral representation
\begin{multline} \label{GenFunk}
     \langle \textrm{e}^{\varphi Q (x)}\rangle_{T, \mu}
        = \sum_{n=0}^\infty \frac{1}{(n!)^2}
	  \biggl[ \prod_{j=1}^n
	   \int\displaylimits_\mathbb{R} \frac{\D p_j}{2 \pi} \,
	   \frac{\E^{\I p_j x}}{1+\E^{\varepsilon(p_j)/T}}
           \int\displaylimits_{\mathbb{R}+\I 0} \frac{\D q_j}{2\pi} \,
	   \E^{-\I q_j x}
	  \biggr]\\[.5\baselineskip]
          \biggl[
	    \prod_{j,k=1}^n \frac{p_j - q_k -\I c}{q_j - q_k -\I c}
	  \biggr]
          \operatorname{det} \bigl[ M (p_j,q_k) \bigr]_{j, k = 1, \dots n}
          \operatorname{det} \bigl[ G (p_j,q_k) \bigr]_{j, k = 1, \dots n}
\end{multline}
for the generating function in the thermodynamic limit by considering
a special scaling limit of the XXZ chain close to the isotropic
ferromagnetic point $\Delta = -1$. Here $M(p,q)$ is defined
as\footnote{Note a typo in equation (26) of the published version of
\cite{SBGK07a}: the first factor of the second term on the rhs should
read $c {\rm e}^\varphi/(w_j - p_k)(w_j - p_k - \I c)$
instead of $c {\rm e}^\varphi/(w_j - p_k)(w_j - p_k + \I c)$.}
\begin{equation} \label{MMatrix}
     M (p,q) =
        \frac{\I}{(p - q)} \biggl[
        \frac{\I c}{p - q + \I c}
	  \prod_{l=1}^n\frac{p - q_l + \I c} {p - p_l + \I c}
        + \frac{\I c \, \E^\varphi}{p - q - \I c} 
	  \prod_{l=1}^n\frac{p - q_l - \I c} {p - p_l - \I c} \biggr]\, .
\end{equation}

The temperature and the chemical potential enter through the functions
$\varepsilon (p)$ and $G(p,q)$ which must be calculated as solutions
of integral equations. The `dressed energy function' $\varepsilon(p)$
is the solution of the non-linear integral equation 
\begin{equation} \label{auxneu}
     \varepsilon(p) = p^2 - \mu
                        - T \int_{\mathbb{R}} \frac{\D q}{\pi} \,
			   \frac{c}{(p - q)^2 + c^2} \,
			   \ln \big(1 + \E^{-\varepsilon(q)/T}\big)
\end{equation}
of Yang and Yang \cite{YaYa69}. The `density function' $G(p,q)$
solves the linear integral equation
\begin{equation} \label{Dichte}
     G (p,q) = - \frac{c}{(p - q)(p - q - \I c)} 
     + \int\displaylimits_{\mathbb{R}}
       \frac{\D k}{\pi} \,
       \frac{c}{(p - k)^2 + c^2} \,
       \frac{G(k,q)}{1 + \E^{\varepsilon(k)/T}} \, .
\end{equation}


\section{Impenetrable Boson Limit} \label{sechardcore}
It is known since the work of Girardeau \cite{Girardeau60} that in the
limit $c \to \infty$ the wave functions of Bosons with contact
interaction turn into those of free Fermions, up to a function which
takes values $\pm 1$. We therefore expect the multiple integral
formula \eqref{GenFunk} to simplify in this limit. In the following
we indicate the limit by supplying a superscript $(\infty)$ to the
respective functions.

We first of all note that the kernel in the integral equations
\eqref{auxneu} and \eqref{Dichte} for $\varepsilon (p)$ and $G(p,q)$
vanishes for $c \to \infty$, such that their solutions become explicit,
\begin{equation} \label{epsGinf}
     \varepsilon^{(\infty)} (p) = p^2 - \mu \, , \quad
     G^{(\infty)} (p,q) = \frac{- \I}{p - q} \, .
\end{equation}
Consequently the expression $1/(1+\E^{\varepsilon(p)/T})$ turns into
the Fermi function
\begin{equation}
     f(p) = \frac{1}{1 + \E^{(p^2 - \mu)/T}}
\end{equation}
for non-relativistic particles. Moreover, the function $M(p,q)$ simplifies
drastically for $c \rightarrow \infty$,
\begin{equation} \label{Minf}
     M^{(\infty)} (p,q) = \frac{\I (1-\E^\varphi)}{p - q} \, ,
\end{equation}
and the explicit product of the right hand side of equation
(\ref{GenFunk}) converges to one.

Substituting (\ref{epsGinf}) and (\ref{Minf}) into the determinants
in (\ref{GenFunk}) we see that they are of Cauchy-type in the limit.
Using the well-known formula
\begin{equation} \label{Cauchy}
     \Det \bigg[\frac{1}{p_j- q_k}\bigg]_{j,k = 1, \dots, n}
        = \frac{\prod_{a < b}{(p_a - p_b)(q_b - q_a)}}
	       {\prod_{a,b=1}^n (p_a - q_b)}
\end{equation} 
we obtain
\begin{multline} \label{simp1}
     \langle \textrm{e}^{\varphi Q(x)}\rangle_{T, \mu}^{(\infty)} = \\
          \sum_{n=0}^\infty \frac{(1-\E^\varphi)^n}{(n!)^2}
          \biggl[ \prod_{j=1}^n \int\displaylimits_\mathbb{R}
                  \frac{\D p_j}{2\pi}\E^{\I p_j x} f(p_j)
                  \int\displaylimits_{\mathbb{R}+\I 0}
		  \frac{\D q_j}{2\pi}\E^{-\I q_j x}
          \biggr]
          \frac{\Delta^2(p)\,\Delta^2(q)}
	       {\prod_{j,k = 1}^n (q_j - p_k)^2} \, ,
\end{multline}
where we introduced the notation
\begin{equation}
     \Delta(p) = \prod_{j < k}(p_k - p_j) = 
        \Det \big[ p_k^{j - 1} \big]_{j,k = 1, \dots, n}
\end{equation}
for Vandermonde determinants.

Due to the symmetry of the integrand with respect to all $q_j$,
one of the Vandermonde determinants $\Delta(q)$ in each term can
be replaced by a product of the diagonal elements of the Vandermonde
matrix, and subsequently the $q$ integrals can be pulled into the
second Vandermonde determinant.\footnote{The same trick was used
for the free Fermion limit of the XXZ chain in \cite{KMST02c}.} Then
\begin{equation} \label{detrep}
     \frac{1}{n!}
     \biggl[ \prod_{l=1}^n
      \int\displaylimits_{\mathbb{R} + \I 0} \mspace{-4.mu}
      \frac{\D q_l}{2\pi} \, \E^{-\I q_l x}
     \biggr]
     \frac{\Delta^2(q)}{\prod_{j,k = 1}^n (q_j - p_k)^2} =
     \Det \biggl[\int\displaylimits_{\mathbb{R} + \I 0} \mspace{-4.mu}
     \frac{\D q}{2\pi} \frac{\E^{-\I q x} q^{j+k-2}}
                            {\prod_{l=1}^n(q - p_l)^2}
     \biggr]_{j,k = 1, \dots, n} \mspace{-6.mu} .
\end{equation}
Obviously the integrals inside the determinant on the right hand
side can now be calculated by means of the residue theorem. Yet,
it turns out to be more convenient to perform a number of elementary
row- and column operations first, resulting in the sequence of identities
\begin{align} \label{seq}
     \Det \biggl[\int\displaylimits_{\mathbb{R} + \I 0}
     \frac{\D q}{2\pi} & \frac{\E^{-\I q x} q^{j+k-2}}
                              {\prod_{l=1}^n (q - p_l)^2}
     \biggr]_{j,k = 1, \dots, n} \notag \\[1ex]
     & = \Det \bigg[
        \int\displaylimits_{\mathbb{R}+ \I 0}
	\frac{\D q}{2\pi} 
        \frac{\E^{-\I q x} \prod_{a=1}^{j-1}(q-p_a)
	      \prod_{b=1}^{k-1}(q-p_b)}{\prod_{l=1}^n(q-p_l)^2}
     \bigg]_{j,k = 1, \dots, n} \notag \\[1ex] 
     & = \frac{1}{\Delta^2(p)} \,
         \Det \bigg[\int\displaylimits_{\mathbb{R}+ \I 0}
	 \frac{\D q}{2\pi}
	 \frac{\E^{-\I q x}}{(q-p_j)(q-p_k)}
     \bigg]_{j,k = 1, \dots, n} \notag \\[1ex]
     & = \frac{1}{\Delta^2(p)} \,
         \Det \bigg[ - \frac{2 \sin(\frac{p_j-p_k}{2}x)}{p_j-p_k} \,
	                        \E^{-\I (p_j+p_k) x/2}
     \bigg]_{j,k = 1, \dots, n} \, .
\end{align}
In the last line for $j = k$ it is understood to take the analytic
continuation of the function
\begin{equation}
     V(u,v) = \frac{2 \sin(\frac{u-v}{2}x)}{u-v} \, , 
\end{equation}
namely $V(u,u) = x$, which comes from the calculation of the residua
at the second order poles of the diagonal entries. Finally, pulling
out the exponential factors and a minus sign in (\ref{seq}), we end
up with the expression
\begin{equation} \label{detintcalc}
     \Det \biggl[\int\displaylimits_{\mathbb{R} + \I 0}
     \frac{\D q}{2\pi} \frac{\E^{-\I q x} q^{j+k-2}}
                            {\prod_{l=1}^n (q - p_l)^2}
     \biggr]_{j,k = 1, \dots, n} =
     \frac{\E^{- \I \sum_{j=1}^n p_j x}}{(-1)^n \Delta^2 (p)}
     \Det \bigl[ V(p_j, p_k) \bigr]_{j,k = 1, \dots, n} \, .
\end{equation}
We further introduce the notation
\begin{equation}
     V_F (u,v) = \sqrt{f(u)} V(u,v) \sqrt{f(v)}
\end{equation}
and substitute it together with (\ref{detrep}) and (\ref{detintcalc})
into (\ref{simp1}) to obtain
\begin{equation} \label{fred}
     \langle \textrm{e}^{\varphi Q (x)} \rangle_{T, \mu}^{(\infty)}
        = \sum_{n=0}^\infty \frac{(\E^\varphi - 1)^n}{n!}
          \Biggl[ \prod_{j=1}^n \int\displaylimits_\mathbb{R}
                  \frac{\D p_j}{2\pi}
          \Biggr] \Det \bigl[ V_F (p_j, p_k) \bigr]_{j,k = 1, \dots, n} \, .
\end{equation}

Now recall (see e.g.\ \cite{WhWa63}) that for an integral operator
$\widehat{K}$ with kernel $K(p,q)$ acting on a function $\phi$ defined
on an interval $I$ as
\begin{equation}
     (\widehat{K} \phi) (p) = \int_I \D q \, K(p,q) \phi(q)
\end{equation}
its Fredholm determinant has the infinite series representation
\begin{equation}
     \det(1 + \widehat{K}) =
          \sum_{n=0}^\infty
	  \frac{1}{n!}
	  \int_{I^n} \D p_1 \ldots \D p_n 
	  \Det \bigl[ K(p_j,p_k) \bigr]_{j,k = 1, \dots, n} \, .
\end{equation}
Then
\begin{equation}
     \langle \textrm{e}^{\varphi Q (x)} \rangle_{T, \mu}^{(\infty)} =
        \det \bigl(1+ {\textstyle \frac{\E^\varphi-1}{2\pi}}
                                           \widehat{V}_F \bigr) \, ,
\end{equation}
where $\widehat{V}_F$ is the integral operator with kernel $V_F (p,q)$.
This is a known result \cite{KBIBo}. What we find remarkable, however,
is that it directly follows from the multiple integral representation
(\ref{GenFunk}).


\section{Free Boson Limit} \label{secsoftcore}
We first of all note that the limit $c\to0$ in \eqref{GenFunk} exists.
In the following we indicate it by a superscript $(0)$ to the
respective functions. Observing that, in this limit, the kernel
in the integral equations \eqref{auxneu} and \eqref{Dichte} for
$\varepsilon (p)$ and $G(p,q)$ turns into a representation of the
$\delta$-function, their solutions become explicit, 
\begin{equation} \label{epsGNull}
     1 + \exp\Big[\frac{\varepsilon^{(0)} (p)}{T}\Big]
        = \exp\Big[\frac{p^2 - \mu}{T}\Big] \, , \quad
	  \frac{G^{(0)}(p,q)}{-c}
        = \frac{1}{(p - q)^2}  \frac{1}{1-\E^{-(p^2-\mu)/T}} \, ,
\end{equation}
and the matrix (\ref{MMatrix}) simplifies,
\begin{equation} \label{MMatrixNull}
     \frac{M^{(0)}(p_j,q_k)}{\I} = \frac{1-\E^\varphi}{(p_j-q_k)^2} \,
    \bigg[\frac{\prod_{a=1}^n (p_j-q_a) }
               {\prod_{a\not=j}^n(p_j-p_a)}\bigg] \, .
\end{equation}

Substituting (\ref{epsGNull}) and (\ref{MMatrixNull}) into the integral
representation \eqref{GenFunk} for $c\to 0$ and using (\ref{Cauchy})
we obtain
\begin{multline} \label{Vorstufe}
     \langle \E^{\varphi Q (x)}\rangle_{T, \mu}^{(0)} = 
        \sum_{n=0}^\infty \frac{(1-\E^{\varphi})^n}{(n!)^2}
        \biggr[\prod_{j=1}^n \int\displaylimits_\mathbb{R}
        \frac{\D p_j}{2\pi}\frac{\E^{\I p_j x}}{\E^{(p_j^2-\mu)/T}-1}
        \int\displaylimits_{\mathbb{R} + \I 0}
	\frac{\D q_j}{2\pi}\E^{-\I q_j x} \biggl] \\
        \det\biggl[\frac{1}{(p_j - q_k)^2}\biggr]_{j,k=1,\ldots,n}^2
        \det\bigg[\frac{1}{p_j - q_k}\bigg]_{j,k=1,\ldots,n}^{-2} \, .
\end{multline}
To simplify the integrand under the $q$ integrals we utilize an identity
borrowed from \cite{KMST02a},
\begin{multline} \label{nicelemma}
     \bigg[\prod_{a<b}(q_b-q_a)\bigg]
     \bigg[\prod_{\substack{a, b = 1 \\ a \not= b}}^n
           (p_a - p_b + \I c) \bigg] \sum_{\sigma\in \mathfrak{S}_n}
          \sign(\sigma) I(p_{\sigma(j)},q_k)\\
        =\bigg[\prod_{a,b=1}^n (p_a-q_b)(p_a - q_b  + \I c) \bigg]
         \det\biggr[
	     \frac{1}{(p_j-q_k)(p_j-q_k+\I c)}\biggl]_{j,k=1,\ldots,n}\, ,
\end{multline}
where the summation extends over all permutations $\sigma \in
{\mathfrak{S}_n}$ and where
\begin{equation}
     I(p_j,q_k) = \frac{\prod_{a=1}^n \big[\prod_{b=1}^{a-1}(p_a-q_b+\I c)
                    \prod_{b=a+1}^n(p_a-q_b)\big]}
		   {\prod_{a<b}(p_b-p_a + \I c)} \, .
\end{equation}
The arguments $p_j$, $q_k$ are understood to represent the sets
$\{p_j\}$, $\{q_k\}$. Performing the free Boson limit $c \to 0$ in
(\ref{nicelemma}) we obtain a useful expression for the quotient
of determinants under the integral in (\ref{Vorstufe}),
\begin{equation} \label{niceaswell}
     \det\bigg[\frac{1}{(p_j-q_k)^2}\bigg]_{j,k=1,\ldots,n}
        \det\bigg[\frac{1}{p_j - q_k}\bigg]_{j,k=1,\ldots,n}^{-1} =
        \sum_{\sigma\in \mathfrak{S}_n}
	\frac{1}{\prod_{a=1}^n(p_{\sigma(a)}-q_a)} \, .
\end{equation}

Squaring of (\ref{niceaswell}) produces two independent sums over
permutations. Then, proceeding similarly as in the case $c \rightarrow
\infty$, we may replace
\begin{equation}
     \bigg[\sum_{\sigma\in \mathfrak{S}_n}
        \frac{1}{\prod_{a=1}^n(p_{\sigma(a)}-q_a)}\bigg]^2 \to
        \frac{n!}{\prod_{a=1}^n(p_{a}-q_a)}
        \sum_{\sigma\in \mathfrak{S}_n}
	\frac{1}{\prod_{a=1}^n(p_{\sigma(a)}-q_a)}
\end{equation}
under the multiple-$q$ integrals due to their symmetry. Consequentially,
the $q$~integrals factorize and can be evaluated by means of the
residue theorem as in (\ref{seq}),
\begin{equation} \label{qInt}
     \int\displaylimits_{\mathbb{R}+\I0}
        \frac{\D q}{2\pi} \frac{\E^{-\I q x}}{(u-q)(v-q)} =
        -V(u,v)\,\E^{-\I(u+v)x/2}  \, .
\end{equation}

Finally, introducing the shorthand notation
\begin{equation}
     b(p) = \frac{1}{\E^{(p^2-\mu)/T} - 1} \, , \quad
     V_B(u,v) = \sqrt{b(u)} V(u,v) \sqrt{b(v)}
\end{equation}
the multiple integral representation of the one-parameter generating
function of the correlation function in the free Boson limit reads
\begin{equation} \label{perm}
     \langle \textrm{e}^{\varphi Q (x)} \rangle_{T, \mu}^{(0)}
        = \sum_{n=0}^\infty \frac{(\E^\varphi - 1)^n}{n!}
          \Biggl[ \prod_{j=1}^n \int\displaylimits_\mathbb{R}
                  \frac{\D p_j}{2\pi}
          \Biggr] \perm \bigl[ V_B(p_j, p_k) \bigr]_{j,k = 1, \dots, n}
	  \, .
\end{equation}
Here the chemical potential must be restricted to the physical range
$\mu<0$ for the integrals involving the Bose function $b(p)$ to converge.
Equation (\ref{perm}) looks rather similar to our former expression
(\ref{fred}) for impenetrable Bosons. The determinants are replaced
by permanents\footnote{The permanent of an $n\times n$ matrix
$A=(A^j_{\phantom{x}k})_{j,k=1,\ldots,n}$ is defined by
\[
     \sum_{\sigma\in \mathfrak{S}_n}
        A^1_{\phantom{x}\sigma(1)}\ldots A^n_{\phantom{x}\sigma(n)} =
        \perm \big[A^j_{\phantom{x}k}\big]_{j,k=1,\ldots,n} \, .
\]
} and the Fermi function is replaced by the Bose function here.


\section{One- and Two-Point Functions}\label{sec12PF}
In this section we would like to recall briefly how to obtain the one- and
two-point density correlation functions by applying (\ref{applgenfun})
to (\ref{fred}) and (\ref{perm}). Because $\varphi$ appears therein only
in the factor $(\E^\varphi-1)^n$, the derivatives with respect to
$\varphi$ can be easily calculated by means of the formula
\begin{equation}
     \partial_\varphi (\E^\varphi-1)^n\Big|_{\varphi=0}
        = \delta_{n, 1} \, , \quad
     \partial^2_\varphi (\E^\varphi-1)^n\Big|_{\varphi=0}
        = \delta_{n, 1} + 2 \delta_{n,2} \, .
\end{equation}
Then we obtain for the impenetrable Boson case the known \cite{KBIBo} explicit expression
\begin{equation}
     \langle j(x) \rangle_{T, \mu}^{(\infty)}
        = \partial_x \int\displaylimits_\mathbb{R}
	  \frac{\D p}{2\pi} f(p) x
	= \int\displaylimits_\mathbb{R}
	  \frac{\D p}{2\pi} f(p) = D_F(T, \mu) \, ,
\end{equation}
which due to translational invariance is independent of the interval
length $x$. The analogous well-known expression for free Bosons,
\begin{equation}
\langle j(x) \rangle_{T, \mu}^{(0)} = 
\int\displaylimits_\mathbb{R} \frac{\D p}{2\pi} \, b(p) = D_B(T,\mu) \, ,
\end{equation}
can be obtained from (\ref{perm}) by application of (\ref{applgenfun}).

As the term arising from $n=1$ is linear in $x$ we need to
consider only the $n=2$ term in order to calculate the density-density
correlation function. For the impenetrable Boson case we have
\begin{align}\notag
     2\frac{\partial^2}{\partial x^2}\frac{1}{2!} &
      \int\displaylimits_\mathbb{R} \frac{\D p_1}{2\pi}
      \int\displaylimits_\mathbb{R} \frac{\D p_2}{2\pi}
      \Det \bigl[  V_F(p_j, p_k) \bigr]_{j,k = 1, 2} \\ \notag
     & = \frac{\partial^2}{\partial x^2}
         \int\displaylimits_\mathbb{R} \frac{\D p_1}{2\pi}
	 \int\displaylimits_\mathbb{R} \frac{\D p_2}{2\pi}\,
	 f(p_1)\, f(p_2) 
      \biggl[
         x^2 - \frac{4 \sin^2(\frac{p_1 - p_2}{2}x)}{(p_1 - p_2)^2}
      \biggr]\\ \notag
     & = 2 \bigg[
           \int\displaylimits_\mathbb{R} \frac{\D p}{2\pi}\,
	   f(p)\bigg]^2 - 2
	   \int\displaylimits_\mathbb{R} \frac{\D p_1}{2\pi}
	   \int\displaylimits_\mathbb{R} \frac{\D p_2}{2\pi}\,
	   f(p_1)\, f(p_2) \cos(p_1 x - p_2 x)\\
     & = 2 D_F^2(T, \mu) - 2
      \biggl[
         \int\displaylimits_\mathbb{R} \frac{\D p}{2\pi}\,
	 f(p) \cos(p x)
      \biggr]^2 \, ,
\end{align}
implying another well-known result \cite{KBIBo},
\begin{equation}\label{2pointoo}
     \langle j(0) j(x)\rangle_{T, \mu}^{(\infty)} - D_F^2(T, \mu)
        = - \bigg[
	  \int\displaylimits_\mathbb{R} \frac{\D p}{2\pi}\,
	  f(p)\, \E^{\I p x}
	  \bigg]^2 \, ,
\end{equation}
for the density-density correlations. Similarly, the
corresponding expression for free Bosons can be obtained from
(\ref{applgenfun}) and (\ref{perm}),
\begin{equation}
\langle j(0) j(x) \rangle_{T, \mu}^{(0)} - D_B^2(T,\mu) =
+ \bigg[
\int\displaylimits_\mathbb{R} \frac{\D p}{2\pi}\, 
b(p)\, \E^{\I p x}
\bigg]^2 \, .
\end{equation}


\section{Conclusion} \label{secconcl}
In the limit $c \rightarrow \infty$ of impenetrable Bosons we
transformed a multiple integral representation for a certain generating
function of the density-density correlation functions of the Bose
gas with contact interaction into a Fredholm determinant of an integrable
integral operator. The latter provides a link to classical integrable
evolution equations and thus a means for a rigorous analysis
of the asymptotics of correlation functions \cite{IIK90b}.
Taking this into account, it is tempting to interpret the multiple
integral representation as a `deformation of a Fredholm determinant'.
One may hope that such an interpretation will finally help
to solve the problem of the calculation of the long-distance
asymptotics also in the generic case of finite repulsion.

For the free Bose gas, $c \rightarrow 0$, we rearranged the summands
in the multiple integral representation in terms of permanents.
Interestingly the resulting series has some similarity with
the Fredholm determinant of the impenetrable case. The Fermi
function is replaced by the Bose function, and the determinants are
replaced by permanents.

In the generic case of finite $c > 0$ our original integral
representation (\ref{GenFunk}) interpolates between the two limiting
cases (\ref{fred}) and (\ref{perm}). What is special about the
limiting cases is, that only the first two terms in the infinite
series contribute to the density-density correlation function.
This property seems lost in the generic case. We expect, however,
that taking into account only a few terms of the series may provide
good approximations for small and for strong finite coupling.


{\bf Acknowledgement.}  AS gratefully acknowledges financial support
by the German Science Foundation under grant number Se~1742~1-1.

\bibliographystyle{amsplain}


\providecommand{\bysame}{\leavevmode\hbox to3em{\hrulefill}\thinspace}
\providecommand{\MR}{\relax\ifhmode\unskip\space\fi MR }
\providecommand{\MRhref}[2]{%
  \href{http://www.ams.org/mathscinet-getitem?mr=#1}{#2}
}
\providecommand{\href}[2]{#2}


\end{document}